\documentclass[12pt]{amsart}

\usepackage{amsfonts}

\begin{document}

\newcommand{\Aut}{{\rm Aut}}
\newcommand{\gr}{{\Gamma}}
\newcommand{\grb}{{W\Gamma}}
\newcommand{\0}{{\varepsilon}}
\newcommand{\VV}{{\rm Ver}}
\newcommand{\hb}{{\hat{b}}}
\newcommand{\half}{{\textstyle{\frac{1}{2}}}}
\newcommand{\li}{{\rm Li}}
\newcommand{\modulo}{{\rm mod}}
\newcommand{\fP}{{\mathcal P}}
\newcommand{\R}{{\mathbb R}}
\newcommand{\SP}{{\rm SP}}
\newcommand{\lag}{{\mathcal L}}

\title {Some measure theory on stacks of networks}\bigskip

\author{Jack Morava}

\address{Department of Mathematics, Johns Hopkins University,
Baltimore, Maryland 21218}

\email{jack@math.jhu.edu}

\thanks{The author was supported by DARPA and the NSF}

\date {15 July 2007}

\begin{abstract} By counting symmetries of graphs carefully
(or equivalently, by regarding moduli spaces of graphs as 
zero-dimensional orbifolds), certain measures on these collections
(elsewhere called `exponential random graphs') can be reinterpreted,
with the aid of special cases of Wick's theorem, as Feynman-style 
measures on the real line. Analytic properties of the latter measures 
can then be studied in terms of phase transitions -- in particular, in 
models for spaces of scale-free trees.
\end{abstract}

\maketitle

\noindent
{\bf Introduction:} This paper is an attempt at an essentially elementary account of a 
single example, which provides evidence for the emergence of an interesting statistical
mechanics of networks. \bigskip

\noindent
For more than fifty years physicists have elaborated techniques introduced by Feynman, which
organize calculations of probabilities on spaces of particle histories as sums over graphs. More
recently [2], researchers have begun to use these methods in reverse, to attack problems in
enumerative combinatorics by analytical means. \bigskip

\noindent
Here [\S 2.3] such (relatively familiar) ideas are formulated as an analog of the
Fourier transform, as a way of evaluating sums over certain spaces of graphs
(known in the literature [15] as exponential random measures) as integrals over the 
real line, with respect to certain Feynman/Gibbs-style measures. Collections of graphs
or networks seem intuitively very discrete, but the transformed sums have interesting
(and accessible) analytic properties, which can sometimes be described in ways similar to the
phase transitions studied in thermodynamics. \bigskip

\noindent
The first section below reviews standard background material, but it contains a significant 
technical point: sums over graphs are really integrals over zero-dimensional moduli spaces, and if
we treat orbifold points respectfully (ie by weighting their symmetries appropriately, by what
physicists call Faddeev - Popov determinants) we get better formulas. Similar issues appear in
other apparently discrete contexts; for example, in Siegel's mass formula for quadratic lattices
[18 VII \S 6.5 (remark)].
\bigskip

\noindent
Without compelling examples this would be empty formalism. Section three sketches 
the basic properties of a model (developed by physicists [3,4,12] interested in (among
other things) the statistics of long-chain polymers) for a phase change in an ensemble of
scale-free trees (ie having nodes of valence $n$ occurring with power-law probability). This
model has some remarkable features: its phase transition is of high (ie fourth) order, and its
critical exponents depend to some extent on the parameters of the model. I hope this note will
help make it, and its many generalizations, accessible for the closer study it deserves. \bigskip  

\noindent
I'd like to acknowledge helpful correspondence and discussions with Zdzislaw Burda, Pawel
Gajer, Soren Galatius, Florin Spinu, and Ben Mann. Special thanks are due to Charles Epstein,
who very kindly straightened out my considerable confusion about the polylogarithm [6]. 
\bigskip

\noindent
This work was supported by DARPA's program on the fundamental questions of biology.
\bigskip

\section{The stack of weighted graphs}
\bigskip

\noindent
This section assembles standard facts and notation from graph theory: \bigskip

\noindent
{\bf 1.1.0} The very familiarity of graphs can lead to confusion. I will be concerned with {\bf
finite, abstract} graphs; such a thing can be defined [8] to be a finite set $G$ with an involution
$\sigma$, together with a retraction $t : G \to G^\sigma$ onto the fixed point set of the
involution. Elements of $G^\sigma$ are the {\it vertices} of the graph, while the elements of the
complement $G - G^\sigma$ are said to be its {\it half-edges}; the quotient of this set by the
involution is the set of edges. The {\it valence} of a vertex $v \in G^\sigma$ is the cardinality
$k(v)$ of $t^{-1}(v)$, ie the number of its incident edges; I will always assume this is positive.
A vertex of valence one defines a {\it terminal edge}, or {\it leg}, of the graph; the remaining
elements of $G^\sigma$ comprise its set $\VV^0(G)$ of {\it internal nodes}. (Legs will always
be terminal.) \bigskip

\noindent
These definitions are somewhat counter-intuitive but they have some virtue; a good test case is
the problem of enumerating the possible labellings of a graph with one edge and two vertices. 
\bigskip

\noindent
Graphs define a category, the morphisms being maps of sets with involution, compatible with the
specified retractions; but isomorphisms will play a particularly important role in this note.
Abstract graphs have natural geometrical realizations as one-dimensional simplicial complexes;
we will not use them here, but they are important in generalizations in which the graphs map to
interesting configuration spaces [5]. \bigskip

\noindent
{\bf 1.1.1} I will always assume that an abstract graph $G$ carries a specified order, or {\it
labeling}, on its set of legs. I will also consider graphs {\it weighted} by a function $w_G$ from
the internal nodes of $G$ to $\{0,1,2, \dots\}$. I will write $\Aut(G)$ for the group of
isomorphisms of $G$, compatible with its labeling and weight function (which will in general be
suppressed from the notation). $\gr_\bullet$ (resp. $\grb_\bullet$) will denote the stack (or
groupoid, or zero-dimensional orbifold) of abstract (resp. weighted) graphs, subject 
to a stability condition on internal nodes (made precise below), with isomorphisms as maps. 
$|\gr_\bullet|$ and $|\grb_\bullet|$ will be the associated sets of isomorphism classes of objects. 
\bigskip

\noindent
{\bf 1.1.2} It is often easiest for technical purposes to work with connected graphs, and I will
write $\gr$ [resp $\grb$] for the subcategories of such things. The genus, or first Betti number, of
a connected graph is $g(G) = E - V + 1$, with $E$ its total number of edges, and $V$ the number
of vertices; this is additive under disjoint union, but for some purposes the Euler characteristic
$\chi(G) \; ( = 1 - g(G)$ for connected graphs) which is also additive, is more useful. \bigskip

\noindent
The number $\nu(G)$ of (external) legs, and the number $\0 = E - \nu$ of internal edges, are
other  useful additive functions. The Euler characteristic (or genus), together with the number of
legs, define a kind of bigrading on the category of graphs. More generally, the generalization 
\[
g_w(G) := \sum_{v \in \VV^0(G)} w_G(v)  +  g(G) 
\]
of the genus is additive on weighted graphs, and defines a similar bigrading on $\grb_\bullet$. 
\bigskip

\noindent
A connected $G$ is {\bf stable} with respect to its weight function, if $2(w_G(v) - 1) + k(v) > 0$ 
for each internal node [10 \S 2]; note that for unweighted graphs (ie with $w_G = 0$) this
precludes nodes of valence two. I'll write $|\gr(\chi,\nu)|$ for the set of isomorphism classes of
connected graphs of Euler characteristic $\chi$ with $\nu$ legs, and so forth. \bigskip

\noindent
There are many other interesting naturally defined functions on these sets, such as the degree 
sequence
\[
D(G) = \{\nu = d_1, d_2, \dots, d_m \}
\]
which assigns to $G$, the partition of $2E$ with $d_k$ equal to the number of vertices of
valence $k$. The distribution of weights, or numbers of legs, define similar functions on
$|\grb(\chi,\nu)|$. \bigskip

\noindent
The function which sends $G$, weighted or not, to its group $\Aut(G)$ of automorphisms is
another important example, as is the related group of automorphisms which are allowed to
change the labelings on the legs. When $G$ is connected, this is just the product of
$\Aut(G)$ with the symmetric group of permutations of the labels, but if not, things are more
complicated; see below. Note that (labeled) trees have {\bf no} automorphisms. \bigskip

\noindent
{\bf 1.2} Graphs in general can be described in terms of the symmetric product of collections
of connected graphs, but we will need to keep careful track of symmetries. \bigskip

\noindent
{\bf 1.2.0} An ordered $n$-tuple of elements from a set $X$ is an element of the $n$-fold
Cartesian product $X^n$ of $X$ with itself. Permuting the order of these elements defines an
action
\[
\Sigma^n \times X^n \to X^n 
\]
of the symmetric group on this set, and the quotient $X^n/\Sigma_n$ is the set of {\bf
un}ordered $n$-tuples of elements from $X$. The coproduct, or disjoint union,
\[
\coprod_{n \geq 0} X^n/\Sigma_n \; := \: \SP^\infty (X)
\]
of these quotients is the free abelian monoid generated by $X$: its elements can be written as
formal finite sums
\[
\sum n_i \{x_i \}
\]
in which the $n_i$ are non-negative integers, and the $x_i$ are (not necessarily distinct)
elements of $X$; of course these formal sums are subject to various rules, such as $n\{ x \} + \{ x
\} = (n+1)\{ x \}$, etc. $\SP^\infty(X)$ is naturally graded by the degree $\sum n_i \{x_i \}
\mapsto \sum n_i$. \bigskip

\noindent
{\bf 1.2.1} The isomorphism class of a general element $G$ of $\gr_\bullet$ can thus be written
as a formal sum $\sum n_i[G_i]$, defined by the disjoint union of $n_i$ copies of (the
isomorphism classes of) connected graphs $G_i$. In other words,
\[
|\gr_\bullet| \; \cong \; \SP^\infty |\gr| 
\]
as graded sets, with the understanding that we now admit a `vacuum' graph with no vertices, cf
eg [4 Ch 7]. If, in the presentation of $G$ as a sum, the indexed components are distinct, then its 
automorphism group  
\[
\Aut(G) \; \cong \; \prod ( \Sigma_{n_i} \wr \Aut(G_i) ) \;.
\]  
is a kind of wreath product. If $\nu = \sum n_i \nu_i = \sum k r_k$, where $r_k$ is the number 
of components of $G$ with precisely $k$ terminal legs, then the group of automorphisms of $G$ 
which are allowed to permute those legs will be a semidirect product of the restricted 
automorphism group with $\prod \Sigma_k^{r_k}$. I'll write 
\[
m(G,\nu) = \frac{\nu!}{\prod k!^{r_k}}
\]
for the number of ways of labeling the legs of $G$. \bigskip

\noindent
{\bf 1.2.2} Finally, it may be worth noting (since the symmetric product construction is not
much used in analysis) that a measure on $X$ pushes forward to define a measure on
$\SP^\infty(X)$.  A measure on connected (weighted) graphs thus extends naturally to define a
measure on general (weighted) graphs. \bigskip

\noindent
The `measures' of most interest in this paper will, however, take values in some field
$\R((\kappa))$ of formal series; $\kappa$ will play the role of a parameter like Planck's constant
in some asymptotic expansion. In fact the formulas will usually include even more parameters. 
\bigskip

\section{A kind of Feynman transform} 
\bigskip       
         
\noindent             
This note is concerned with measures on spaces of (weighted) graphs, which is a subject
with a large and somewhat disconnected literature; in [15 \S 5, 16] the term `exponential random
graphs' is used for a class of examples somewhat wider than those which will be in focus here.
\bigskip

\noindent                                 
{\bf 2.1} That general class of models employs a family $\epsilon_k$ of functions on
(isomorphism classes of) graphs, together with corresponding parameters $\beta_k$, called
`inverse temperatures'; a graph $G$ is then assigned probability 
\[
P_\beta(G) = \exp(-\sum \beta_k \epsilon_k(G))
\]
(perhaps suitably normalized). If the $\epsilon$'s are additive (under disjoint union of graphs),
these probabilities will be multiplicative, and their values on general graphs will be extended
from their values on connected graphs as described in the preceding section. \bigskip

\noindent
That will be the case here, with some modifications: $\fP_\beta$ will be defined by functions
$\epsilon_{w,k}(G)$ which count the total number of vertices in $G$ with weight $w$ and
valence $k>1$, and I will also introduce parameters $t, \lambda$ and $\kappa$ to keep track of
the number of legs, edges and Euler characteristic (which are all of course additive). The
literature of exponential random graphs allows more complicated $\epsilon$'s, which count more
general subconfigurations (various kinds of polygons, etc) of $G$; such data can be incorporated
into measures of the form $F(G) \fP_\beta(G)$. \bigskip

\noindent
In this paper the key difference involves the enumeration of graphs with symmetries. If we define
\[
\fP_\beta(G) \: = \: \frac{m(G,\nu)}{ |\Aut(G)|} \exp(-\sum \beta_{w,k}
\epsilon_{w,k}(G))
\]
then our `measure' $\mu_{\beta,\lambda}$ (actually taking values in the formal power series ring 
$\R((\kappa))[[t,\lambda]]$) will be defined by summing the function
\[
G \mapsto \fP_\beta(G) \kappa^{-\chi(G)}\lambda^{\0(G)} \frac{t^{\nu(G)}}{\nu(G)!}
\]
over subsets of $|\grb_\bullet|$. The extra combinatorial factors in the definition of $\fP_\beta$
divide by the order of the group of extended automorphisms of the graph (allowed  to permute
the labels on the legs). Note that the order $|\Aut(G)|$ is not quite multiplicative under disjoint
union. \bigskip

\noindent
{\bf 2.2} The interest of this class of measures is that under certain circumstances they
can be calculated, or at least described, in terms of formal measures defined by {\bf local} 
interactions in a one-dimensional Euclidean space. \bigskip

\noindent
Feynman introduced certain expressions of the general form
\[
\int f(x) \; \exp(-\kappa^{-1} \lag(x)) \; dx \;,
\]
interpreted as perturbed Gaussian integrals over infinite-dimensional spaces, and he showed that
these could be in some sense evaluated in terms of  formal series involving more and more
complicated but finite-dimensional integrals; but it was the physicist Gian-Carlo Wick who
systematized the combinatorics behind those calculations. This section is concerned with a
drastically simplified case of Wick's result, which provides a rigorous asymptotic expansion for
expressions like the integral above. In this account the domain of integration will be taken to be
one-dimensional [9, 10], but the techniques below generalize quite naturally to `non-linear sigma
models', in which the graphs are mapped by the parameter $x$ to some kind of configuration
space [7].  \bigskip

\noindent
{\bf 2.2.1} In our situation
\[
\lag(x) = \half \lambda^{-1} x^2 - V(x) 
\]
will be an analog of the Lagrangian function of physics. Its first term represents the quadratic
term in the exponent of the background Gaussian integral, while 
\[
V(x) = \sum_{k \geq 0} b_k \frac{x^k}{k!}
\]
plays the role of an `interaction potential'; but we will allow coefficients
\[
b_k = \sum_{g \geq 0} b_{g,k} \kappa^g \in \R[[\kappa]]
\]
which are power series in the asymptotic parameter. However, we need to require that $b_{0,0}
= b_{0,1} = b_{0,2} = b_{1,0} = 0$,  ie that the coefficients $b_{g,k} = 0$ unless $2(g-1) + n >
0$ (or, alternately: $V(0) \equiv 0 \; \modulo \; \kappa^2, \; V'(0) \equiv V''(0) \equiv 0 \; \modulo
\; \kappa)$; this signals a secret connection with moduli spaces of Riemann surfaces [9, 13]. This
extra generality will not be used in the main example in \S 3, but it illustrates some of the
flexibility of this class of models. \bigskip

\noindent
{\bf 2.2.2} Let $Z(0)$ denote $\int_\R \exp(-\kappa^{-1} \lag(x)) \; dx$ (or, more precisely, its
representation as an asymptotic series); then we can define the `expectation value' of a function
$f$ on the line to be
\[
\langle f \rangle \; \sim \; Z(0)^{-1} \int_\R f(x) \; \exp (- \kappa^{-1}\lag(x)) \; dx \;.
\]
If we now write 
\[
\beta_{g,k} = - \log b_{g,k} \;,
\]
then Wick's theorem takes the form of an asymptotic expansion 
\[
\langle \exp(\kappa^{-1}tx) \rangle \; \sim \; e^{\half \kappa^{-1} \lambda t^2}[ \; 1 \; + \; 
\sum_{G \in |\grb_\bullet^+|} \fP_\beta(G) \kappa^{-\chi(G)} \lambda^{\0(G)}
\frac{t^{\nu(G)}}{\nu(G)!} 
\; ] 
\]
(independent of the domain of integration, as long as it contains the origin). The sum is taken
over the set $|\grb_\bullet^+|$ of weighted `non-vacuum' graphs: connected or not, but such that 
each component has at least one leg. From now on the leading term in right-hand expression will 
be included in this sum, on the grounds that it can be regarded as the contribution of the empty 
set, understood as a non-vacuum graph. \bigskip

\noindent
Note that when $V = 0$, this formula reduces to the classical fact that the Fourier-Laplace
transform of a Gaussian function is again Gaussian. \bigskip

\noindent                                                            
{\bf 2.2.3} We can use some formal version 
\[
\langle f \rangle = \int_\R \hat{f}(\xi) \langle \exp(i \xi x) \rangle \; d \xi 
\]
of Plancherel's theorem to extend this result. Since the asymptotic expression depends only on
the germ of $f$ at the origin, we will regard $f \in \R[[x]]$ as a power series; substituting $t = i
\kappa \xi$ gives
\[
\langle f \rangle \; \sim \; 2 \pi \sum_{|\grb_\bullet^+|}\partial^\nu \exp(\half \kappa \lambda
\partial^2) \; f|_{x = 0} \cdot \fP_\beta \lambda^\0 \kappa^{-\chi + \nu}/\nu! \;.
\]
If $\gamma_k(x) = x^k/k!$ is the $k$th divided power of $x$, for example,
then the coefficient 
\[
\partial^\nu \exp(\half \kappa \lambda \partial^2) \; \gamma_k(x)|_{x = 0} \; = \; \gamma_m(\half
\kappa \lambda)
\]
if $k-\nu = 2m$ is even, and is zero otherwise. Note that the resulting ``measures'  are really
formal sums of delta-functions and their derivatives, all with support at the origin! \bigskip

\noindent
{\bf 2.2.4} As formulated in [10], Wick's theorem in one dimension asserts that 
\[
(2 \pi \kappa \lambda)^{-1/2}\int_\R \exp \kappa^{-1} \Bigl( tx - \half \lambda^{-1} x^2  + V(x)
\Bigr) \; dx \; \sim \; \exp \kappa^{-1} \Bigl( \half \lambda t^2 \; + \hat{V}(t)\Bigr) \;,
\]
where the formal transform
\[
\hat{V}(t) =  \sum \hb_{\chi,\nu} \; \kappa^g \; \frac{t^\nu}{\nu!}        
\] 
has coefficients
\[
\hb_{\chi,\nu} = \sum_{G \in |\gr (\chi,\nu)|} |\Aut(G)|^{-1} \lambda^{\0(G)} \prod_{v \in 
\VV^0(G)} b_{w(v),k(v)} \;.
\]
The term $\half \lambda t^2$ will sometimes be interpreted as the contribution to $\hat{V}$ 
from the unique graph in $\gr(1,2)$ (whose coefficient is not defined by the prescription
above). \bigskip

\noindent
The expansion in \S 2.2.2 just rewrites this: when $t=0$, 
\[
(2\pi \kappa  \lambda)^{-1/2}\int \exp(-\kappa^{-1} \lag(x)) \; dx \; \sim \; \exp(\kappa^{-1}
\hat{V}(0)) \;,
\]
with $\hat{V}(0) = \sum_{g > 1} \hb_{g,0}\kappa^g$ (since $2(g-1) + n > 0$). This is a sum
over `vacuum graphs' (without legs), so the `renormalized' sum $\hat{V}_+(t)  = \hat{V}(t) -
\hat{V}(0)$ contains contributions only from graphs with at least one leg. Thus
\[
\langle \exp(\kappa^{-1}tx) \rangle = Z(0)^{-1} \int \exp \kappa^{-1}(tx - \lag(x)) \; dx \; \sim \;
\exp\kappa^{-1}(\half \lambda t^2 + \hat{V}_+(t)) \;,
\]
and the formula in \S 2.2 comes from expanding the exponential of $\kappa^{-1}\hat{V}_+$: we
get 
\[
\prod_{|\gr(\chi,\nu)|, \nu \geq 1} \exp(|\Aut|^{-1} P_\beta \kappa^{-\chi} \lambda^\0
\frac{t^\nu}
{\nu!}) \; = \; \sum \prod \frac {[P_\beta(G_i) \kappa^{-\chi(G_i)} \lambda^{\0(G_i)} 
t^{\nu(G_i)}]^{k_i}}{|\Aut(G_i)| k_i! \; \nu(G_i)!^{k_i}}
\]
which, everything being more or less additive, rearranges into 
\[
\sum \frac{\nu(G)!}{|\Aut(G)|\prod \nu(G_i)!^{k_i}} P_\beta(G) \kappa^{-\chi(G)}
\lambda^{\0(G)} \frac{t^{\nu(G)}}{\nu(G)!} = \sum_{[\grb_\bullet^+]} \fP_\beta \kappa^{-\chi}
\lambda^\0 \frac{t^{\nu}}{\nu!}
\]
(where $G = \sum k_i G_i$, with the $G_i$ distinct). \bigskip

\noindent
{\bf 2.3} We can summarize this elementary calculation as follows: \medskip

\noindent
We consider two formal measure spaces, \medskip

$\bullet$ the real line $\R$, with respect to the formal measure 
\[
\mu_\lag \sim (2 \pi \kappa \lambda)^{-1/2} \int_\R \exp(-\kappa^{-1} \lag(x)) \; dx \;,
\]

\noindent
(supported at 0, as noted above), and 
\medskip

$\bullet$ $|\grb_\bullet|$, with respect to the measure $\mu_{\beta,\lambda}$ defined by
summing $\fP_\beta \lambda^\0 \kappa^{-\chi+ \nu}/\nu!$. 
\medskip

\noindent
Let $X_\nu$ be the characteristic function of the set of non-vacuum weighted graphs with
exactly $\nu$ legs, and define 
\[
\Phi_k = \sum_{m \geq 0} \gamma_m(\half \kappa \lambda) X_{k-2m} \;.
\]
With this notation, we have the \medskip

\noindent
{\bf Theorem:} The linear operator
\[
f = \sum f_k \gamma_k \mapsto \check{f} = \sum f_k \Phi_k
\] 
maps the space $L^1(\R,\mu_\lag)$ of formal functions isometrically to the formal span of the
$X_\nu$ in $L^1(|\grb_\bullet|,\mu_{\beta,\lambda})$. \bigskip

\noindent
This correspondence sees functions on the line only in the neighborhood of zero, and their
transforms lie in the subspace spanned by the characteristic functions $X_\nu$; but we can
nevertheless probe the space of graphs in more detail, by varying the parameters $\lambda$ and
$\beta$. \bigskip

\section{Ensembles of scale-free trees} \bigskip

\noindent
This general framework seems very rich, even after rather drastic specialization. This section is
an account of some measures on spaces of trees defined by potentials              
 \[
V(x) = \sum_{k \geq 3} \frac{x^k}{k^{\alpha+1}} \; \; \; \; (= \li_{\alpha+1}(x) - x -
2^{-\alpha-1} x^2) 
\]
of polylogarithmic form. [The `missing' linear and quadratic terms are tracked by the parameters
$t$ and $\lambda^{-1}$; the conventions used here seem to simplify book-keeping in the long
run [9].] The parameter $\kappa$ does not appear in this formula, so we will be concerned with
{\bf un}weighted graphs; in fact we will be mostly concerned with trees (ie graphs of genus zero,
which however will not be rooted). \bigskip

\noindent
This subject has a rather extensive literature in condensed-matter physics, closely related to the
study of `scale-free' networks [3,4,12], but one of the purposes of this note is to argue
that it deserves more attention from mathematicians. In particular, it seems to me that this model
is potentially at least as rich as (for example) the Ising model. It may also be useful to note that
related path integral techniques have found other interesting applications in mathematical
biology, eg [17]. \bigskip

\noindent
{\bf 3.1} The partition function for weighted graphs is the power-series expansion of an analytic 
function
\[
Z(t,\lambda) = (2 \pi \kappa \lambda)^{-1/2} \int_\R \exp( - \kappa^{-1}(tx - \lag(x))) \; dx 
\]
near $t = \lambda = 0$. Following ideas pioneered by Landau [1, 14 Ch VI], we can try to
interpret its singularities in terms of phase transitions. \bigskip

\noindent
The leading term in the statistical free energy for this ensemble of networks is the tree
approximation
\[
W(t) = \lim_{\kappa \to 0} \kappa \log Z \; \sim \;  \half \lambda t^2 + \lim_{\kappa \to 0} 
\hat{V}(t) = \sum_{G \in |\gr(1,\nu)|, \nu \geq 2} P_\beta(G) \lambda^\0 \frac{t^\nu}{\nu!} \;;
\]
the last equality uses the fact that labeled trees have no automorphisms. We can omit the factorial
in the denominator of the last term by writing it as a sum over isomorphism classes of {\bf
un}labeled trees. \bigskip

\noindent
{\bf 3.2} A formal analog of Laplace's method for evaluating integrals with large parameter
(ancestral to the saddle point or `stationary phase' approximation), at a critical point $x_0$ of 
\[
\phi(x) = \phi(x_0) + \half \phi'' (x_0) \cdot (x - x_0)^2 + \dots \;,
\]
is an expression of the form
\[
(2\pi \kappa)^{-1/2} \int_\R \exp(- \kappa^{-1} \phi(x)) \; dx = \frac{\exp(-\kappa^{-1}
\phi(x_0))}{{\phi''(x_0)}^{1/2}} \; [1 + {\rm higher \; order \; in} \; \kappa \dots ] \;.
\]
This implies a formula 
\[
W(t,\lambda) = tx_0 - \lag(x_0) 
\]
for the tree level free energy, interpreted as a function of $t,\lambda$ via the equation 
$t = \lag'(x_0)$ for a critical point of $tx - \lag(x)$. Rewriting that equation in iterable form
yields
\[
x_0(t,\lambda)  = \lambda(t + V'(x_0)) = \lambda t + \lambda V'(\lambda t) + \dots \in
\R[[t,\lambda]] \;;
\]
more generally, the implicit function theorem guarantees the existence of a solution $x =
x_0(t,\lambda)$ off the critical locus $\lambda^{-1} = V''(x_0)$, ie away from  
\[
t_0 = x_0 V''(x_0) - V'(x_0) \;.
\] 
\medskip

\noindent
{\bf 3.3} However, the polylogarithm has a subtle singularity at $x = 1$: 
\[
\li_{\alpha+1}(e^{-x}) \; \equiv\; \Gamma(-\alpha)(x \sim i0)^\alpha
\]
modulo smooth functions, where
\[
(x \sim i0)^\alpha := \half[(x + i0)^\alpha + (x - i0)^\alpha] = x_+^\alpha + \cos \pi \alpha \; x_-
^\alpha  
\]
is the average of two possible Gel'fand-Shilov regularizations of the complex power. [The first
draft of this paper contained a very confusing sign error here; I'm very sorry about that. A fuller
discussion of this formula has been since posted at [6 Prop. 3].]  Thus, if the potential has the
form specified at the beginning of this section, then at $x=1$ we recover equation (26)
\[
t_0 = V''(1) - V'(1) = 1 + \zeta(\alpha - 1) - 2 \zeta(\alpha) 
\]
of [3] for the locus of singularities in the model. 
\bigskip

\noindent
As $\alpha$ varies, this defines an approximate hyperbola in the positive quadrant of the
$(t,\alpha)$ plane, partitioning it into two regions, as in Fig. 6 of [4]. Left of this line, the
resulting measure is expected to be concentrated on trees with an exponential degree distribution,
while to its right it is thought to be supported on `clumpy' concentrations, with many edges
abutting on a relatively small number of hubs. The critical line defines a phase transition, with
the measure concentrated on the {\bf scale-free} trees. \bigskip

\noindent
The behavior of $x_0$ as a function of $\lambda$ near the critical line varies, depending on
whether $\alpha$ is greater or less than two. If $\alpha > 3$, the first coefficient
\[
\frac{\partial \lambda}{\partial x_0} = - x_0^{-2}(t + V'(x_0)) + x_0^{-1}V''(x_0)
\]
in the Taylor series for $\lambda = \lambda(x_0)$ goes to zero near the critical line, so  
\[
x_0 = 1 - C(\lambda_0 - \lambda)^{1/2} + \cdots \;.
\]
On the other hand the saddle-point equation
\[
\lambda^{-1} = x_0^{-1}(t + V'(x_0))
\]
can be rewritten
\[
\lambda^{-1} = \lambda_0^{-1} + x_0^{-1}(t + V'(x_0)) - (t_0 + V'(1)) \;,
\]
(with $\lambda_0 = V' (1)$); as $t \to t_0$ this specializes to 
\[
\lambda^{-1} = \lambda_0^{-1} + (x_0^{-1} - 1)V' (1) + x_0^{-1}(V'(x_0) - V'(1)) \;.
\]
When $3 > \alpha > 2$, 
\[
\lambda^{-1} \sim \lambda_0^{-1} + \Gamma(1 - \alpha)(1 - x_0)^{\alpha - 1} + \cdots \;,
\]
so 
\[
x_0 \sim 1 - C(\lambda_0 - \lambda)^{1/(\alpha - 1)} + \cdots \;.
\]
It is tempting to take
\[
\eta = \log x_0 \sim (\lambda_0 - \lambda)^{1/(\alpha - 1)} 
\]
(when $\alpha \in (2,3)$) as an order parameter for this transition [1,19 I \S 3]. \bigskip

\noindent
Some further results about the higher-genus case can be found in [12]; I hope to return to these,
and other questions about the generalized thermodynamical properties of these models, soon.
\bigskip

\newpage

\bibliographystyle{amsplain}

(2002) 
\end{document}